\documentclass[iop]{emulateapj}
\shortauthors{Brook et al.}
\submitted{Recieved 19 Nov. 2013}

\newcommand{\hMpc}{{\ifmmode{h^{-1}{\rm Mpc}}\else{$h^{-1}$Mpc}\fi}}
\newcommand{\hkpc}{{\ifmmode{h^{-1}{\rm kpc}}\else{$h^{-1}$kpc}\fi}}
\newcommand{\hMsun}{{\ifmmode{h^{-1}{\rm {M_{\odot}}}}\else{$h^{-1}{\rm{M_{\odot}}}$}\fi}}
\newcommand{\ltsima}{$\; \buildrel < \over \sim \;$}
\newcommand{\gtsima}{$\; \buildrel > \over \sim \;$}
\newcommand{\lsim}{\lower.5ex\hbox{\ltsima}}
\newcommand{\gsim}{\lower.5ex\hbox{\gtsima}}

\def\lesssim{\mathrel{\hbox{\rlap{\hbox{\lower4pt\hbox{$\sim$}}}\hbox{$<$}}}}
\def\gtrsim{\mathrel{\hbox{\rlap{\hbox{\lower4pt\hbox{$\sim$}}}\hbox{$>$}}}}

\newcommand{\beq}{\begin{equation}}
\newcommand{\eeq}{\end{equation}}
\def\beqa{\begin{eqnarray}}
\def\eeqa{\end{eqnarray}}
\def\hMpc{$h^{-1}\,{\rm Mpc}$}
\def\hkpc{$h^{-1}\,{\rm kpc}$}

\def\Msun{$M_\odot$}
\def\Mhalo{$M_{\rm halo}$}

\def\MSMHR{$M_*$$-$$M_{\rm halo}$}

\begin{document}
\title{The stellar-to-halo mass relation for  Local Group galaxies}


\author
       {C. B. Brook$^{*,}$\altaffilmark{1},
        A. Di Cintio\altaffilmark{1,2},
         A. Knebe\altaffilmark{1}, 
         S. Gottl\"{o}ber\altaffilmark{3},
          Y. Hoffman\altaffilmark{4},
          G. Yepes\altaffilmark{1} ,          
          S. Garrison-Kimmel\altaffilmark{5} 
          }
\affil{$^1$Departamento de F\'isica Te\'orica, M\'odulo C-15, Facultad de Ciencias, Universidad Aut\'onoma de Madrid, 28049, Madrid, Spain
}
\affil{$^2$Physics Department G. Marconi, Universit\`{a} di Roma Sapienza, Ple Aldo Moro 2, 00185 Rome, Italy}
\affil{$^3$Leibniz Institute for Astrophysics Potsdam, An der Sternwarte 16, 14482 Potsdam, Germany}
\affil{$^4$Racah	Inst.	of	Physics	Hebrew	University	Jerusalem	91904,	Israel}
\affil{$^5$  Center for Cosmology, Department of Physics and Astronomy, University of California, Irvine, CA 92697, USA}

\begin{abstract}
We contend that a single power law halo mass distribution is appropriate for direct matching to the stellar masses of observed Local Group dwarf galaxies, allowing the determination of the slope of the stellar mass-halo mass relation for low mass galaxies.  Errors in halo masses are well defined as the Poisson noise of simulated local group realisations, which we determine using  local volume simulations.
For the stellar mass range $10^7$\Msun$<$$M_*$$<$$10^8$\Msun, for which we likely have a complete census of observed galaxies,  we find that  the stellar mass-halo mass relation  follows a power law with slope of 3.1, significantly steeper than most values in the literature. 
 This steep  relation between stellar and halo masses would  indicate that Local Group dwarf galaxies are hosted by dark matter halos with a  small range of mass.  Our methodology is robust down to the stellar mass to which the census of observed Local Group galaxies is   complete,  but the significant  uncertainty in the currently measured slope of the stellar-to halo mass relation will decrease dramatically if the Local Group completeness limit was  $10^{6.5}$\Msun\ or below,
highlighting the importance of pushing such limit  to lower masses and larger volumes.
\end{abstract}


\section{Introduction} \label{sec:introduction}
By comparing stellar masses of galaxies from large scale surveys with  masses of halos in cosmological dark matter  simulations, one can use abundance matching techniques to derive the stellar-to-halo mass relation, \MSMHR \, \citep[e.g.][]{moster10,guo10}. More direct measurements of   \MSMHR \, can also be made by measuring halo masses using, for example, galaxy-galaxy lensing \cite[e.g.][]{hoekstra04,hudson13} or satellite dynamics \cite[e.g.][]{prada03,more11}, with the various methods giving reasonable agreement \cite[e.g.][]{leauthaud12}.

However, the range of masses that can be probed by abundance matching is limited by the luminosity down to which galaxy surveys are complete, and by   variations in the halo mass functions  of simulations. Large scale galaxy surveys, e.g. SDSS and GAMMA,  have provided complete  stellar mass functions \citep{baldry08,baldry12}  down to $\sim 10^8$\Msun within volumes that are large enough such that the mass function within collisionless cosmological simulations  have  variations which are insignificant.

The details at the low mass end,  $M_*$$\lsim$$10^9$\Msun, become less clear, as does the question as to how to extend the relation  to even  lower mass galaxies. For example, the stellar mass function from \cite{baldry08} has an upturn at the low mass end,  which translates to an upturn in the \MSMHR\ relation \citep{behroozi13}. This implies a slope of  the \MSMHR\  relation of  $\alpha$$=$$1.6$ at the low mass end, where $M_*$$\propto$\Mhalo$^\alpha$.   Extraploating this relation to lower masses would imply that  low stellar mass galaxies  would reside in significantly lower mass dark matter halos than predicted by extrapolating the earlier models of \cite{moster10} and \cite{guo10}, which found steeper slopes (higher values of $\alpha$) for the \MSMHR\ relation.  

However, \cite{garrison-kimmel14} point out that extrapolating a slope of $\alpha$$=$$1.6$  would significantly over-estimate the number of Local Group galaxies with $M_*\gsim$5$\times10^{6}$\Msun . Using updated observational data from \cite{baldry12}, which shows less upturn in the stellar mass function and hence a steeper \MSMHR\ relation,   \cite{garrison-kimmel14} find a slope $\alpha$$=$$1.92$ at the low mass end.

Regardless of these differences, there is no {\it a piori} reason to believe that the relation between stellar mass and halo mass should be extrapolated to  low mass galaxies, $M_*$$<$$10^{8}$\Msun. Further,  the relation derived from large volume galaxy surveys may not be directly applicable to the particular environment of the Local Group. 

In this study, we use constrained simulations of the local universe (CLUES) to show that the mass function of a volume analogous to  the Local Group follows a single power law. This allows us to match the masses of dark matter halos taken from the underlying power law mass function directly to
 the stellar masses of observed Local Group galaxies. We thus provide the first robust measurement of the relation between the stellar mass of Local Group galaxies and the masses of the halos in which they are hosted, assuming a $\Lambda$CDM cosmology. 

\section{Local Group Simulations} \label{sec:simulation}

All simulations in this paper are dark matter only. 
Within the  CLUES\footnote{www.clues-project.org $\qquad ^*$Email: cbabrook@gmail.com} project, the Hoffman-Ribak algorithm \citep{hoffman91} is used to generate  initial conditions as constrained realizations of Gaussian random fields using observational data of the local environment. A series of  realizations were run in order to obtain  a local group candidate with Milky Way and Andromeda (MW/M31) analogue halos with proper masses, relative positions and with negative radial velocity. 

CLUES have been extensively used for other investigations  \citep[e.g.][]{Gottloeber10,Libeskind10,Knebe11a,dicintio13a}. We repeat here only the basic information. Our fiducial run  has WMAP5 cosmology \citep{wmap5}, i.e. $\Omega_m$$=$$0.28$, $\Omega_{\Lambda}$$=$$0.72$ and $h$$=$$0.7$, a normalization of $\sigma_8 $$=$$0.8$ and power spectrum slope of $n$$=$$0.96$.  The local group volume has been simulated within a cosmological box with side length of $L_{\rm box}$$=$64$h^{-1}$Mpc  using \texttt{GADGET2} \citep{Springel05}.  The  mass resolution is $m_{\rm  DM}$$=$$3.6$$\times$$10^{5}$\Msun\   and the gravitational softening length is $\epsilon$$=$$137$pc. The two most massive halos, taken as analogues of the MW and M31, have virial masses of  $M_{\rm MW}$$=$$1.7\times10^{12}$\Msun\ and $M_{\rm M31}$$=$$2.2\times10^{12}$\Msun\ and are separated by a distance of 849\,kpc. 

For calculating $rms$ errors between simulated halo masses and the underlying power law mass function, we also use a CLUES local group run with WMAP3 cosmology, and 10 simulated local group volumes from  the Exploration of the Local Group Volume In Simulation Suite (ELVIS, \citealt[][]{garrison-kimmel14}). 

The \texttt{AHF} halo finder\footnote{http://popia.ft.uam.es/AHF} \citep{Gill04b,Knollmann09} has been used to identify all (sub-)halos in our simulation. Virial mass is defined as the mass within a sphere containing $\Delta_{vir}\simeq350$ times the cosmic mean matter density.  We measure halo masses  at their maximum mass, prior to being stripped.  

\begin{figure}
\hspace{-.1cm}\includegraphics{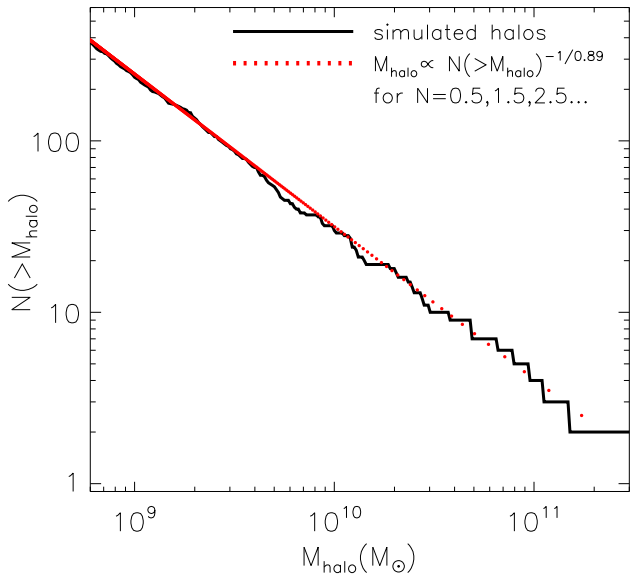}
 \caption{The mass function of  CLUES halos, including sub-halos,  within the defined local group region (LGV, black line).
 The red dots follow the best fit to the halo mass function,  with masses assigned to each N= 0.5, 1.5, 2.5 etc. by inverting the mass function to give \Mhalo$\propto$ N($>$\Mhalo)$^{-1/0.89}$.    }
\label{fig:massfn}
\end{figure}

\section{Results}\label{sec:results}

In what follows, the slope of the dark matter halo mass function,  $\alpha_{\rm dm}$, is defined via a power law fit to the mass function, N($>$\Mhalo)\,$\propto$\Mhalo$^{\alpha_{\rm dm}}$. The slope, $\alpha$, of the \MSMHR\ is defined by assuming a power law fit   $M_*$$\propto$\Mhalo$^{\alpha}$. The  local group volume (LGV) is defined as a sphere of radius 1.8\,Mpc centred on the MW analogue halo.

Figure~\ref{fig:massfn}  shows (black line)  the halo mass function of the fiducial CLUES local group  simulation within the LGV.   Sub-halos, using their maximum mass values prior to stripping, are included within the total halo population. 
The LGV mass function follows a single power law with  slope $\alpha_{\rm dm}$$=$$-$0.89,  the same slope as the cosmological box from which the LGV is drawn, and  essentially the same slope as other cosmological  mass functions in the literature \cite[e.g.][]{jenkins01}. 

The  LGV of the CLUES simulation run with  WMAP3 cosmology  also has a mass function slope of  $\alpha_{\rm dm}$$=$$-$0.89,  the same as the slope in the cosmological volume from which it is drawn. Further, the 10 LGVs  surrounding paired MW/M31 analogue halos from  from \cite{garrison-kimmel14}, follow a slope $\alpha_{\rm dm}$$=$$-$0.9.  These results provide strong evidence  that the mass function of  Local Group dwarf galaxies is a single power law. 



\begin{figure}
\hspace{-.1cm}\includegraphics{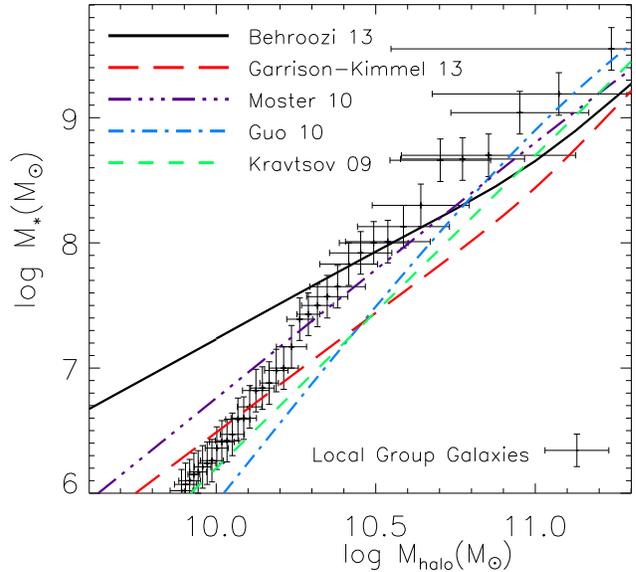}
 \caption{The stellar mass of observed Local Group galaxies within 1.8\,Mpc of the Milky Way,  assigned to halo masses from the power law fit to the halo mass function (red dots in Figure~\ref{fig:massfn}). Stellar mass errors are 0.17$dex$ (Woo et al. 2008). Halo mass errors are the $rms$ deviations of simulated LGV halo masses   from the power law fit to the halo mass function.    Also shown are the stellar-halo mass relations of Behroozi et al. 2013 (solid line), Garrison-Kimmel et al. 2013 (long-dashed line), Moster et al. 2010 (triple-dot-dashed line), Guo et al, 2010 (dot-dashed line),   and the luminosity-mass relation from  Kravtsov  2010 (dashed line).  Scatter in the halo mass-stellar mass would only change the matching order: shifting high stellar mass galaxies into low mass halos must be accompanied by low stellar mass galaxies being hosted by high mass halos.  The slope will not be flattened.  }
\label{fig:MsMh}
\end{figure}

The red dots in Figure~\ref{fig:massfn} are halo masses of a defined local group distribution which follows the mass function power law: masses are assigned to each N= 0.5, 1.5, 2.5 etc.,  according to the inverted mass function, 
\beq
\frac{M_{\rm halo}}{10^{10}M_{\odot}} = N_0\times N(>M_{\rm halo})^{-1.12}
\eeq
\noindent where $N_0$$=$47.9  for our fiducial model, and $N_0$$=$38.1 
for the mean of the 12 simulated LGVs. 

This distribution of masses is appropriate for direct application to the observed Local Group, with the virial masses of each halo subject only to Poisson noise around the power law halo mass function.

We next match this power law halo mass distribution to observed stellar masses of Local Group galaxies, assuming a one-to-one correspondence in order of mass, as shown in Figure~\ref{fig:MsMh}. Our predicted Local Group abundance matching is shown as points with error bars, while stellar-to-halo mass relations from previous studies are shown as lines. Observed Local Group galaxy luminosities  are taken from  \cite{mcconnachie12}, with stellar mass to light ($M_*/L$) taken from \cite{woo08}. Updated distance measurements have resulted in slight changes to the \cite{woo08} stellar masses \citep[see][for a table of updated stellar masses for most galaxies]{kirby13}. We assume $M_*/L$=1.6 for  galaxies not listed in either  \cite{woo08} or \cite{kirby13}. There are 41 galaxies in our sample with $M_*$$>$10$^6$\Msun\ and within 1.8\,Mpc of the Milky Way.

Error bars for stellar masses in Figure~\ref{fig:MsMh} are 0.17\,$dex$, the  quoted typical error in \cite{woo08}. 
Error bars for \Mhalo\ in Figure~\ref{fig:MsMh} are $rms$ errors of each  simulated halo mass from the corresponding (by ordered number) power law halo mass, \Mhalo$\propto$N$^{-1/\alpha}$, using our full suite of 12 simulated LGVs. These errors in halo mass  are dominated by Poisson noise, with other sources of error coming from the different cosmologies and any sample variance being insignificant.  
The errors on the halo masses are reasonably small in the relevant region for this study, where N($>$\Mhalo)$>$10.  This region is relavent because there are 10  Local Group galaxies with M$_*$$\gsim$10$^{8}$\Msun. so the single power law in this region can be used to match galaxies with M$_*$$\lsim$10$^{8}$\Msun. 

\begin{figure}
\hspace{-.1cm}\includegraphics{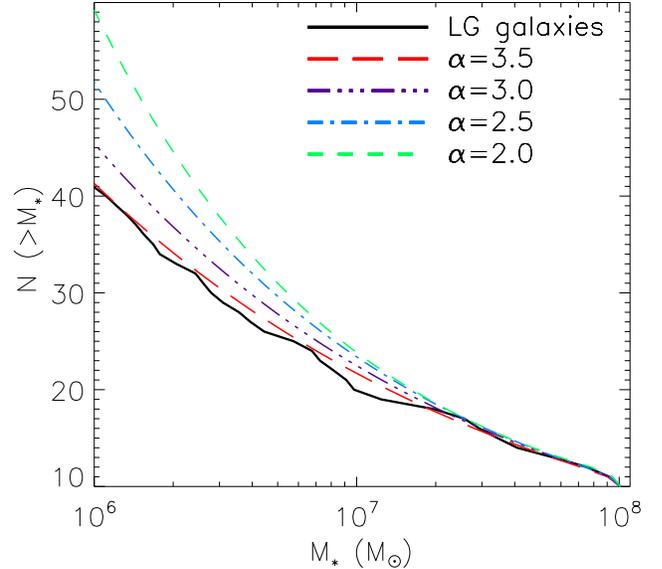}
 \caption{The stellar mass function of observed Local Group galaxies within 1.8\,Mpc of the Milky Way (black line).  Coloured lines are stellar mass functions that assume a halo mass function of slope $\alpha_{\rm dm}$$=$$-$0.89 along with various assumed slopes for a power law \MSMHR\ relation, $\alpha$$=$$2, 2.5, 3, 3.5$ plotted as dashed, dot-dashed, triple-dot-dashed and long-dashed lines  respectively. Normalisation ensures 10 galaxies with $M_*$$>$10$^8$\Msun in each case: we are interested  in the shapes of the curves.}
 \label{fig:MSfn}
\end{figure}
 

Analyzing Figure~\ref{fig:MsMh} we found that the  \MSMHR\ relation for the Local Group galaxies is well fit by $\alpha$$=$3.1 in the region $10^7$$<$$M_*/M_\odot $$\le$$10^8$. 
The catalogue of Local Group galaxies within 1.8\,Mpc of the MW is likely complete down to  $M_*$$=$$10^7$\Msun\ and possibly down to $M_*$$=$5$\times$$10^6$\Msun\ \citep{koposov08,tollerud08}.  If we do increase the assumed completeness range to $5$$\times$10$^6$$<$$M_*/M_\odot$$\leq$$10^8$ we obtain a slope of $\alpha$$=$3.5.

In this context, we note the recent discovery of satellites in the vicinity of M31 that have  stellar masses of several times $10^6$\Msun\ \citep[]{martin13a,martin13b}, demonstrating that we are certainly not complete down to $M_*$ $\sim$10$^6$\Msun\ in the Local Group. 
 Even using the $M_*$ $=$$10^7$\Msun\ limit, our derived slope of the \MSMHR\ relation, $\alpha$$=$3.1, is nevertheless significantly steeper than  most values in the literature (see the Introduction and Figure~\ref{fig:MsMh}). 

The normalisation of the mass function coming from the total mass of the Local Group will not affect such derived value of $\alpha$; instead, the effect will be to shift all points in Figure~\ref{fig:MsMh} left or right. In the region $10^7$$<$$M_*/M_\odot $$\le$$10^8$, the  \MSMHR\ relation  is fit by

\beq
M_{*} = \left( \frac{M_{\rm halo}}{M_0\times10^{6}}\right) ^{3.1}
\eeq
\noindent where $M_0$$=$$79.6$ in our fiducial run. $M_0$$=$$63.1$  when using the  mean power law mass function for the 12 LGVs.
Similarly, systematic errors in observed stellar mass determinations, such as assuming a different initial mass function, will shift all points up or down. 


In Figure~\ref{fig:MSfn}, we plot the stellar mass function of observed Local Group galaxies, shown as the black line. Fixing the halo mass function slope at $\alpha_{\rm dm}$$=$$-0.89$, we examine the resultant stellar mass functions for various assumed values of the slope of the \MSMHR\ relation, $\alpha$. Stellar mass functions that result from  assuming $\alpha$$=$$2, 2.5, 3, 3.5$ are plotted as dashed, dot-dashed, triple-dot-dashed and long-dashed lines respectively  in  Figure~\ref{fig:MSfn}. Normalisation ensures 10 galaxies with $M_*$$>$10$^8$\Msun, i.e. the observed number,  in each case. 


The stellar  mass functions for low values of $\alpha$ diverge from the observed function as we go to low stellar masses.  Down to a  completeness limit of $M_*$$\sim$10$^7$\Msun, values of  $\alpha$ are hard to distinguish, with just a few galaxies separating $\alpha$$=$$2$ from $\alpha$$=$$3.5$.  However,  assuming that the completeness limit for Local Group is closer to 5$\times$10$^{6}$\Msun, a slope of  $\alpha \gsim 3$ is clearly favoured. 
As the catalogue of observed Local Group galaxies becomes complete to lower stellar masses, the stellar mass-halo mass relation will become increasingly well defined.

In Figure~\ref{fig:MSfn}, no assumption  of a one-to-one correspondence between halo mass and stellar mass is made, as it is in Figure~\ref{fig:MsMh}. The slope of the stellar mass function, $\alpha_*$ is simply derived from the relation, 1+$\alpha_*$$=$$(1+\alpha_{dm})/\alpha$. Any scatter around the \MSMHR\ relation,  which may be  large for low mass galaxies \cite[e.g.][]{behroozi13}, will not affect our result for the preferred slope, $\alpha \gsim 3$.



\section{Conclusions}\label{sec:conclusion}

Supported by evidence from constrained local group simulations, we argue that a power law mass function for halos is appropriate to be directly applied to Local Group galaxies. Poisson noise of simulated realizations  provide well defined errors in halo masses. By matching such power law mass function to stellar masses of observed Local Group galaxies, we determine a slope of the \MSMHR\ relation of $\alpha$$=$$3.1$ for galaxies with stellar mass $M_*$$\lsim$10$^{8}$\Msun. This  determination of the relation for Local Group galaxies down to $M_*$$=$10$^{7}$\Msun\ is significantly steeper than most values in the literature, which have generally been extrapolations of the abundance matching relation from higher masses. 

Our value of $\alpha$ is consistent with the extrapolation of the \cite{guo10} relation to small stellar masses, yet the upturn in the relation at masses above $M_*$$=$10$^{8}$\Msun\ \citep{behroozi13,garrison-kimmel14} indicates that the extended relation is likely to be more complex than a single power law for masses $M_*$$<$10$^{9}$\Msun.

The key insight of our paper comes from the fact that the halo mass function of the local group analog simulation in a volume with radius 1.8\,Mpc follows a single power law slope to large enough masses to host all the local group galaxies with $M_{*} \lsim$$10^{8}$\Msun. Because the census of observed Local Group galaxies is well known down to $M$$\sim$10$^7$\Msun\ (or a little lower) within such a volume, we can match Local Group dwarf galaxies to dark matter halos drawn from a population that follows a power law in a region where Poisson noise, which determines the uncertainties in the halo masses, is low.

On the other hand, smaller volumes, such as sub-halo mass functions of  Milky Way analogue halos, are not constrained in this manner. {\it On average}, such populations will also follow a single power law \citep[e.g.][]{boylan-kolchan10a}, but their individual  mass functions  are dominated by Poisson noise. 
Therefore the Local Group is unique, with a volume that is large enough to have a single, well defined power law mass function, yet small enough to find faint galaxies.

Scatter in the \MSMHR\ relation  appears evident in the Local Group. For example, Fornax is 100 times more luminous than Draco, but seems to have a smaller halo mass  \citep{penarrubia08}.  However, such scatter will not flatten the slope of the stellar-to-halo mass relation that we derived, as is evident in  Figure~\ref{fig:MSfn} where only slopes are considered, with no assumption made regarding how stellar masses and halo masses are matched. Nevertheless, scatter  in the  \MSMHR\ relation will result in  some relatively high stellar mass galaxies being hosted by relatively low mass halos,  but this is only achieved in conjunction with relatively low stellar mass galaxies being hosted by high mass halos. 

As we show in Figure~\ref{fig:MSfn}, surveys of Local Group galaxies that extend the the completeness limit to lower luminosities, such as {\tt SKYMAPPER} \citep{keller07} and {\tt LSST} \citep{lsst}  will provide increasingly strong constraints on the \MSMHR\ relation, and particularly on the slope of such relation at small masses, simply by matching observed data to a power law mass function for dark matter halos.


\acknowledgements
We thank Brad Gibson and Alan McConnachie for feedback on an early draft. 
CBB, ADC, AK and GY received support from MINECO (Spain)  grant  AYA2012-31101. AK is supported by MINECO through the Ram\'{o}n y Cajal programme and  grants CSD2009-00064 (MultiDark Consolider project), CAM~S2009/ESP-1496 (ASTROMADRID network).
The CLUES simulations were performed and analyzed at the Leibniz Rechenzentrum Munich (LRZ) and the Barcelona Supercomputing Center (BSC). 
We thank DEISA for  access to  computing resources  through DECI projects SIMU-LU and SIMUGAL-LU.
SG and YH have been partially supported  by  Deutsche Forschungsgemeinschaft  grant $\rm{GO}563/21-1$.
YH has been partially supported by the Israel Science Foundation (1013/12).
GY acknowledges support from the Spanish MINECO under research grant FPA2012-34694 and Consolider Ingenio 
SyeC CSD2007-0050 and from Comunidad de Madrid under ASTROMADRID project(S2009/ESP-1496). 



\end{document}